\documentclass{Interspeech}
\usepackage{array}
\usepackage{float}
\usepackage{graphicx}
\usepackage{xcolor}
\usepackage{booktabs}
\usepackage{wrapfig}
\usepackage{subfig}
\usepackage{cite}
\usepackage{makecell}
\usepackage[justification=centering]{caption}

\interspeechcameraready

\title{Can Emotion Fool Anti-spoofing?}
\author[affiliation={1}]{Aurosweta}{Mahapatra}
\author[affiliation={1}]{Ismail Rasim}{Ulgen}
\author[affiliation={2}]{Abinay}{Reddy Naini}
\author[affiliation={2}]{Carlos}{Busso}
\author[affiliation={1,3}]{Berrak~}{Sisman}

\affiliation{}{Center for Language and Speech Processing, Johns Hopkins University}{USA}
\affiliation{Language Technologies Institute}{Carnegie Mellon University}{USA}
\affiliation{}{Data Science and AI Institute (DSAI), Johns Hopkins University}{USA}

\email{amahapa2@jhu.edu, iulgen1@jhu.edu, anaini@andrew.cmu.edu, busso@cmu.edu, sisman@jhu.edu}
\keywords{Anti-Spoofing, Speech Deepfake, Text-to-Speech, Ensemble Modeling, Emotion}

\usepackage{comment}
 \usepackage{color}
 \usepackage{xcolor}
\begin{document}

\maketitle

\begin{abstract}


Traditional anti-spoofing focuses on models and datasets built on synthetic speech with mostly neutral state, neglecting diverse emotional variations. As a result, their robustness against high-quality, emotionally expressive synthetic speech is uncertain. We address this by introducing EmoSpoof-TTS, a corpus of emotional text-to-speech samples. Our analysis shows existing anti-spoofing models struggle with emotional synthetic speech, exposing risks of emotion-targeted attacks. Even trained on emotional data, the models underperform due to limited focus on emotional aspect and show performance disparities across emotions. This highlights the need for emotion-focused anti-spoofing paradigm in both dataset and methodology. We propose GEM, a gated ensemble of emotion-specialized models with a speech emotion recognition gating network. GEM performs effectively across all emotions and neutral state, improving defenses against spoofing attacks. We release the EmoSpoof-TTS Dataset \footnote{EmoSpoof-TTS Dataset: https://emospoof-tts.github.io/Dataset/}.


\end{abstract}

\section{Introduction}

Anti-spoofing in the speech domain focuses on detecting spoofed speech produced through replay techniques, speech synthesis, and voice conversion \cite{AS_Survey}. Among these, the misuse of synthetic speech \cite{AS_Survey, misinfosurvey} has become a growing concern, especially with advancements in emotionally expressive text-to-speech (TTS) models. 

Recent TTS models can effectively capture a range of emotions, making synthetic speech increasingly realistic ~\cite{Emodiff, ED-TTS, prompttts++, lee2022hierspeech,le2023voicebox,vall-e}. Furthermore, zero-shot TTS models \cite{StyleTTS2,F5TTS,Cosyvoice} have demonstrated the ability to produce realistic synthetic speech from just a few seconds of reference audio while effectively capturing human emotions. While state-of-the-art (SOTA) anti-spoofing models \cite{RawNet2, AASIST} perform relatively well in detecting neutral synthetic speech, their effectiveness against these emotional TTS models remains uncertain. This necessitates robust countermeasures against emotion-targeted attacks, where an attacker exploits the model’s sensitivity to a particular emotion.


\setlength{\parskip}{0.2em}
Various challenges, such as the Automatic Speaker Verification and Spoofing Countermeasures Challenge (ASVspoof) \cite{ASVspoof2015} and the Audio Deep Synthesis Detection (ADD) Challenge \cite{ADD2022, ADD2023}, have been introduced to raise awareness about anti-spoofing. These initiatives have led to the development of large-scale datasets, including ASVspoof 2019 \cite{ASVspoof2019}, ASVspoof 2021 \cite{ASVspoof2021}, and ASVspoof 2024 \cite{ASVspoof2024}, which serve as benchmarks in this field. However, these datasets primarily contain speech in a neutral state and do not specifically address the emotional aspects of speech, leaving a gap in the study of emotion-driven spoofing attacks. 


The ASVspoof and ADD challenges, along with large-scale datasets, have significantly advanced the development of anti-spoofing models. Traditional models \cite{LCNN,ASSERT,ResNet34} rely on handcrafted features for spoof speech detection. However, more recent end-to-end models like RawNet2 \cite{RawNet2} and AASIST \cite{AASIST} process raw audio waveforms directly, achieving highly promising results by eliminating the need for manually engineered features. However, even these SOTA models have primarily been trained and evaluated on traditional datasets \cite{ASVspoof2019, ASVspoof2021, ASVspoof2024}, which focus on neutral speech and lack diverse emotional variations. As a result, current anti-spoofing models are not explicitly designed to account for the impact of emotion on spoof detection, making them vulnerable to emotionally expressive synthetic speech. 


\setlength{\parskip}{0.2em}


In this work, we focus on the neglected emotional aspect of anti-spoofing by exploring whether the emotions can fool anti-spoofing systems. 
We reveal the limitations of current anti-spoofing paradigm in terms of lacking both suitable datasets and emotion tailored methods. To address this limitations, we build an emotionally rich anti-spoofing dataset and propose an anti-spoofing method specifically designed to tackle emotional speech to mitigate the risks associated with the emotion-targeted attacks.


We introduce EmoSpoof-TTS, a diverse dataset containing over 29 hours of emotionally expressive synthetic speech generated using recent TTS models. Our findings reveal that current anti-spoofing models are unreliable in scenarios involving emotional speech and exhibit performance disparities across different emotions. In addition, to improve the performance on emotionally expressive synthetic speech, we propose the Gated Ensemble Method (GEM). GEM leverages multiple emotion-specialized anti-spoofing models with a Speech Emotion Recognition (SER) model, which acts as a gating mechanism to enhance its performance in detecting high-quality, emotionally expressive speech while also addressing performance disparities across emotions. Our key contributions can be summarized as:
\begin{itemize}
    \item We identify and highlight the significant limitations of existing anti-spoofing models on emotionally expressive synthetic speech and expose a new risk of emotion-targeted attacks.
    \item  We introduce and release EmoSpoof-TTS, a corpus of emotionally expressive synthetic speech generated using recent TTS models.
    \item We propose the Gated Ensemble Method (GEM), which significantly improves emotional synthetic speech detection, highlighting the need to prioritize emotion in anti-spoofing design.
\end{itemize}

\section{Related Work}

State-of-the-art anti-spoofing models, such as RawNet2 \cite{RawNet2} and AASIST \cite{AASIST}, are well known for their strong benchmark performance. These models employ an end-to-end architecture, eliminating the need for hand-crafted feature extraction and simplifying implementation. In this study, we select RawNet2 as our baseline anti-spoofing model due to its robust performance. As illustrated in Figure \ref{fig:figure1}, RawNet2 comprises a SincNet layer, residual blocks, a GRU layer, fully connected layers, and an output layer. Most prior studies utilize the publicly available pre-trained RawNet2 model, trained on the ASVspoof 2019 dataset \cite{ASVspoof2019}. We note that as TTS continues to advance, the ASVspoof datasets—despite their scale and popularity—lack emotional speech samples. This limitation raises concerns about whether high-performing anti-spoofing models remain effective against increasingly emotional synthetic speech.


Recently, EmoFake \cite{EmoFake} was introduced as an emotional fake dataset generated via Emotional Voice Conversion (EVC) techniques. While valuable, EmoFake focuses on voice conversion and does not encompass emotional text-to-speech (TTS) synthesis. To bridge this gap, we introduce a new dataset, EmoSpoof-TTS, and utilize it to assess and enhance anti-spoofing robustness against advanced synthesis techniques and emotionally expressive speech. Further details are provided in Section 3.


\section{Anti-spoofing vs Emotion}

The state-of-the-art anti-spoofing models are primarily evaluated on traditional datasets, which lack emotionally expressive speech and do not account for recent advancements in speech synthesis. As a result, they are unreliable with emotional speech, making them vulnerable to emotion-targeted attacks. This section introduces EmoSpoof-TTS and compares anti-spoofing model performance on traditional datasets versus EmoSpoof-TTS. 


\subsection{EmoSpoof-TTS}
We propose EmoSpoof-TTS to explore the impact of emotions in anti-spoofing. This dataset focuses on spoofed speech generated by zero-shot TTS models—StyleTTS2 \cite{StyleTTS2}, F5-TTS \cite{F5TTS}, and CosyVoice \cite{Cosyvoice}—known for producing high-quality, realistic speech even with a short reference audio. These TTS models requires emotional reference speech in order to synthesize emotional speech. In this work, we have utilized The Emotional Speech Database (ESD) \cite{ESD} for this purpose. ESD dataset consists of 10 speakers expressing five emotions (Happiness, Anger, Sadness, Neutral, etc) with 350 utterances per emotion.
For speech synthesis, we separate 50 utterances per speaker and emotion as reference speech. Using these reference speech, we synthesize 300 emotional speech samples which have the same content, speaker identity and emotion with the 300 bona-fide speech samples in ESD per speaker and emotion category.

We consider four basic emotions: \textit{Happiness}, \textit{Anger}, \textit{Sadness}, and \textit{Neutral state}. EmoSpoof-TTS contains a total of 36,000 synthesized speech samples from 4 emotions, 10 (5 male, 5 female) speakers and 3 TTS models\cite{StyleTTS2,F5TTS,Cosyvoice} which is parallel to 12,000 bona-fide samples from ESD. EmoSpoof-TTS featuring multiple emotions, speakers, and TTS models, provides a valuable resource for studying emotional speech in anti-spoofing. We will publicly release the dataset to foster research in this area.




\begin{figure}[t!]
    \centering
\includegraphics[width=0.45\textwidth]{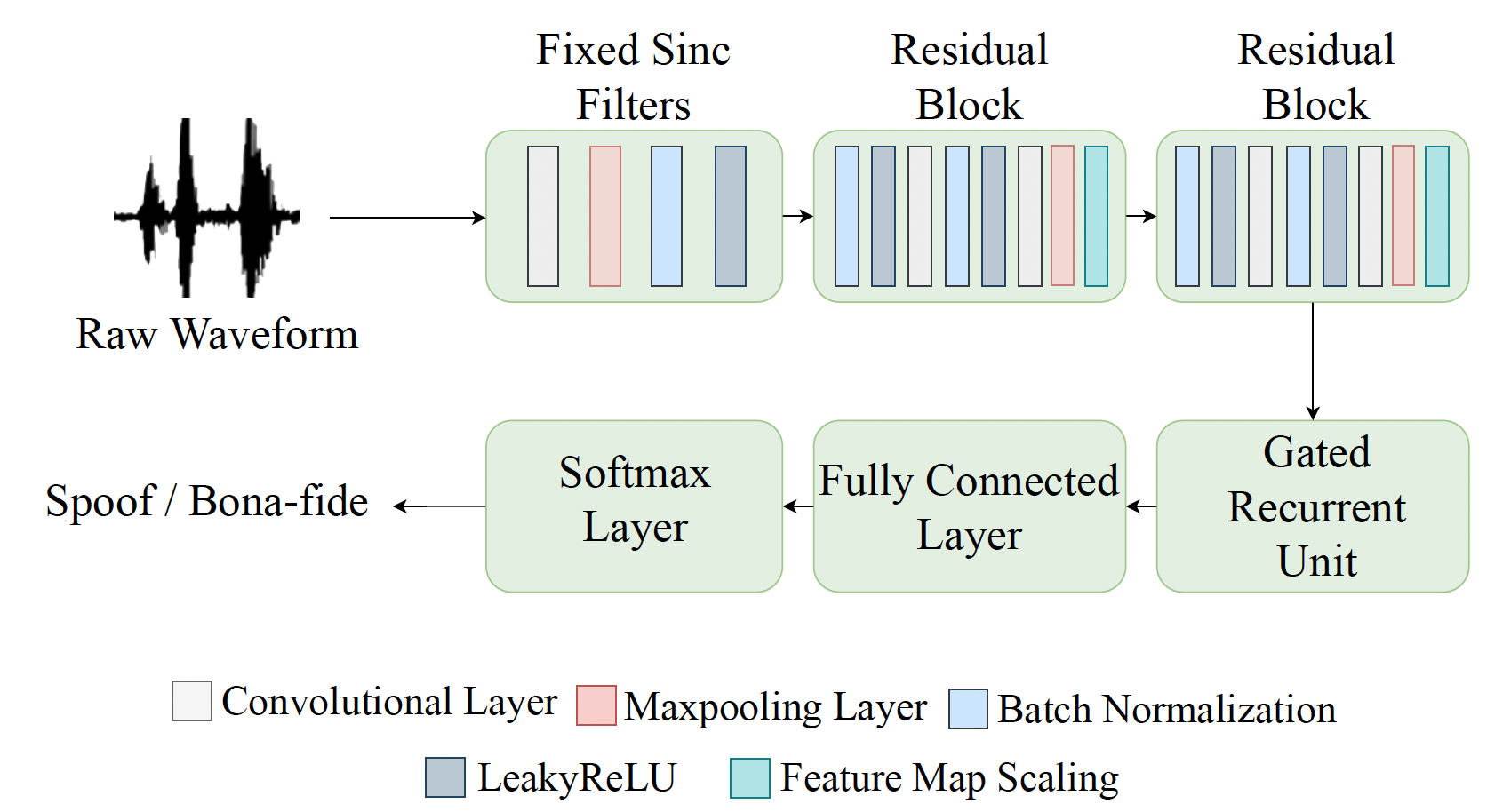} 
    \caption{Architecture of RawNet2}
    \label{fig:figure1}
\vspace{-7mm}
\end{figure}




\begin{figure*}[!th]
    \centering
    \subfloat[Training Phase: Two-Stage Fine-Tuning Process]{
        \includegraphics[width=0.4\textwidth, height=6.0cm]{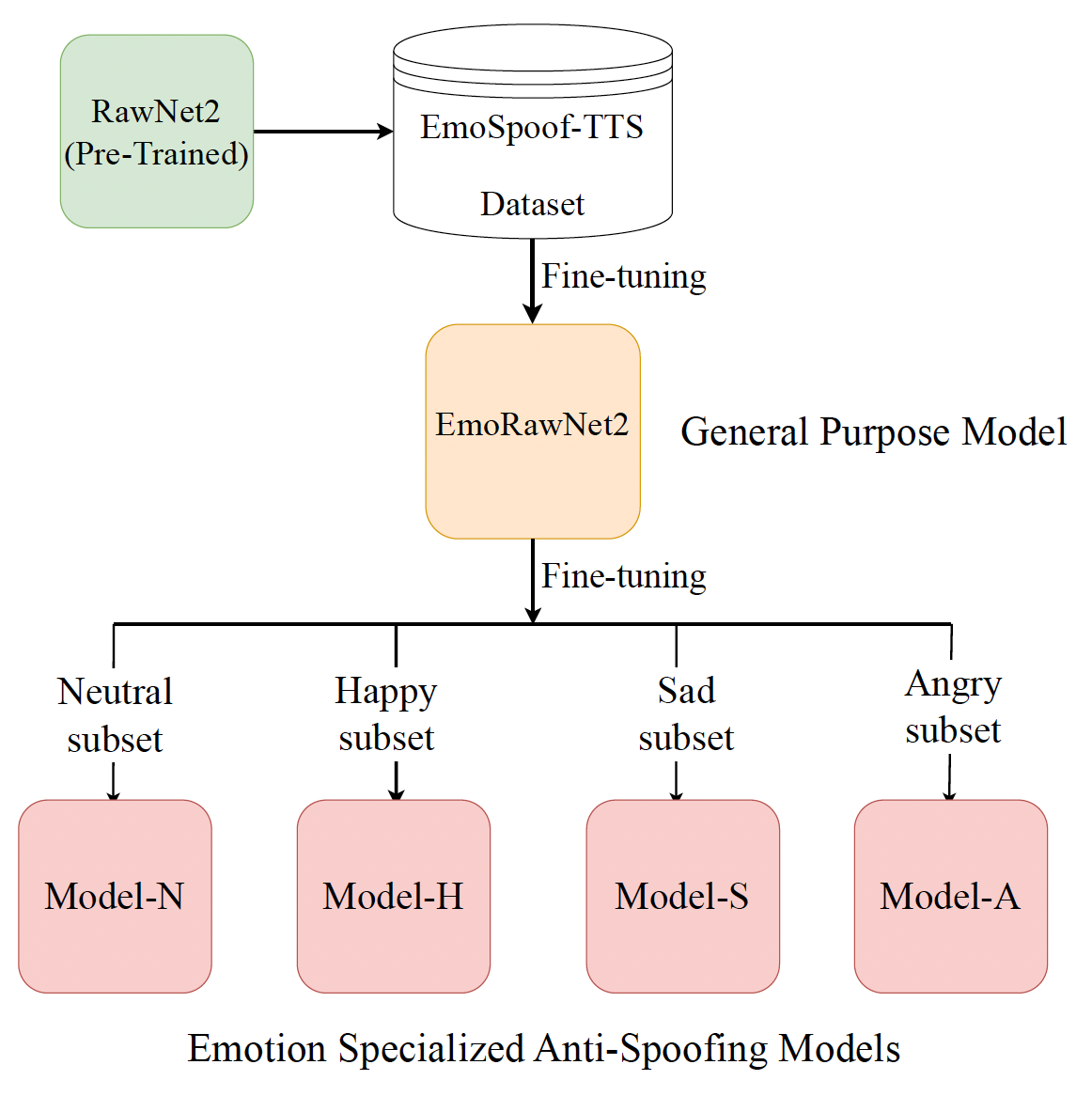}
        \label{fig:subfig1}
    }
    \hfill
    \subfloat[Inference Phase: Computing the Weighted Average Prediction Score]{
        \includegraphics[width=0.5\textwidth, height=6.0cm]{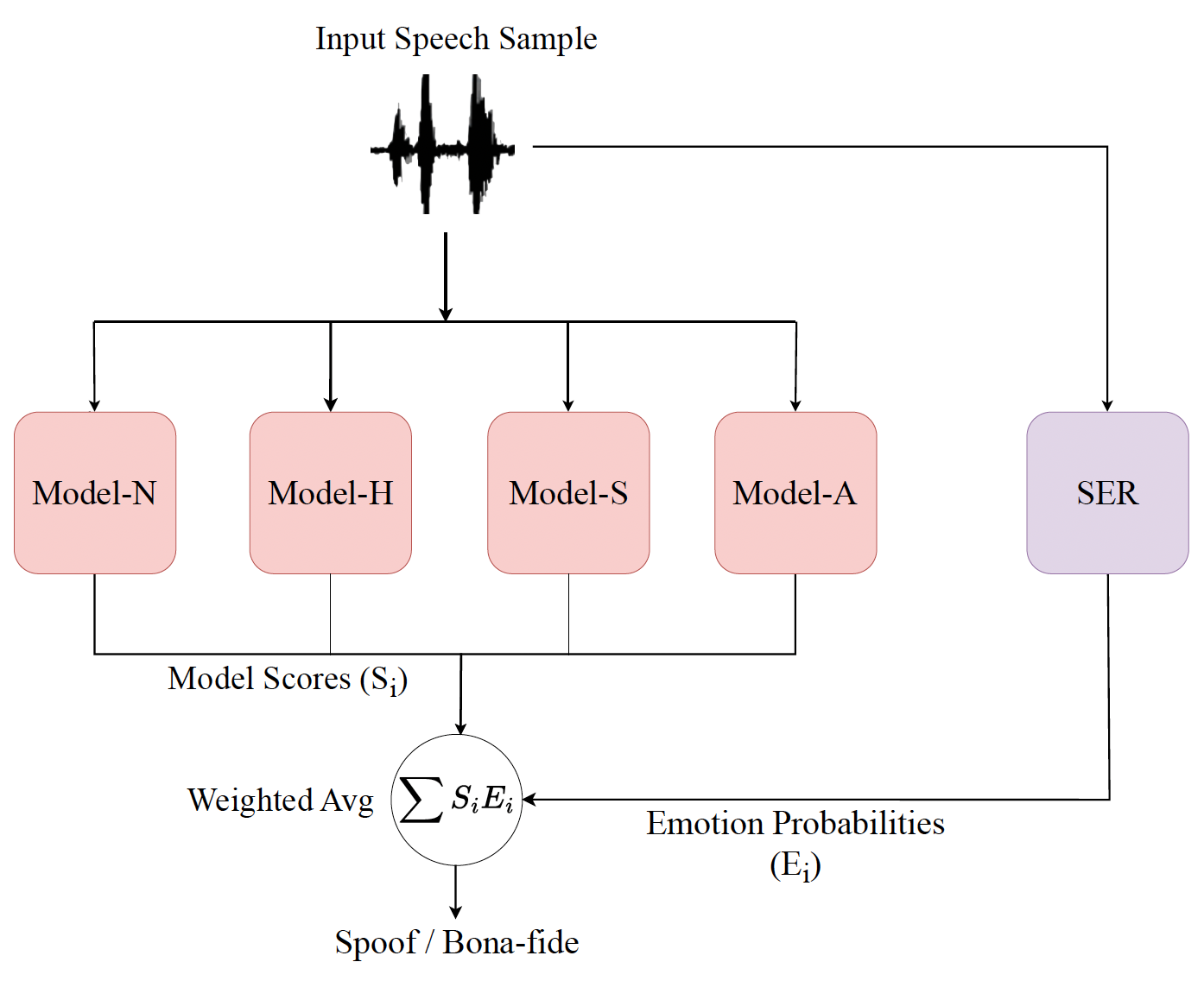}
        \label{fig:subfig2}
    }
    \vspace{-1mm}
    \caption{Proposed Gated Ensemble Method (GEM)}
    \label{fig:figure2}
    \vspace{-3mm}
\end{figure*}

\vspace{-1mm}
\subsection{Analysis}


We evaluate the pre-trained RawNet2 model, trained on the Logical Access (LA) partition of the ASVspoof 2019 dataset, as the traditional anti-spoofing method.

We first assess its performance using evaluation datasets from LA track for ASVspoof 2019, ASVspoof 2021, and the Track 1 for ASVspoof 2024. As shown in Table 1, RawNet2 performs well on ASVspoof 2019 and ASVspoof 2021, with low Equal Error Rate (EER) values. However, its performance drops significantly on ASVspoof 2024, a more recent dataset, highlighting its limitations in generalizing to evolving spoofing techniques.


We then evaluated the pre-trained RawNet2 model on EmoSpoof-TTS, using spoof samples from four speakers while maintaining speaker consistency across all three TTS models. The results, shown in Table \ref{tab:emospoof}, are reported for individual emotions and a combined category (Happiness, Anger, and Sadness, referred to as HAS). The HAS category allows us to assess the model’s performance on emotionally expressive speech compared to neutral speech, while individual emotions highlight detection performance across different emotional states.
\renewcommand{\arraystretch}{1.3}
\setlength{\textfloatsep}{1pt}
\begin{table}[t!]
\centering
\caption{Performance of Pre-Trained RawNet2 model on test-set of Traditional Datasets (EER\%)}
\label{tab:asvspoof}

\scalebox{0.8}{\begin{tabular}{c|c|c|c}
    \hline
      & \multicolumn{3}{c}{\textbf{Test Dataset}} \\ 
    \cline{1-4}
    \textbf{Model} & \textbf{ASVspoof 2019} & \textbf{ASVspoof 2021} & \textbf{ASVspoof 2024} \\ 
    \hline
    \textbf{RawNet2} & 4.60 & 14.73 &  40.67 \\ 
    \hline
\end{tabular}}
\vspace{2mm}
\end{table}


The evaluation presented in Table \ref{tab:emospoof} provides insights into the vulnerabilities of the pre-trained RawNet2 model when exposed to high-quality, emotionally expressive synthetic speech. The model struggles to detect fake samples from EmoSpoof-TTS, with EER values higher than those for the ASVspoof 2024 dataset across all TTS models. The EER for HAS across all three models is consistently higher than for Neutral. For instance, the EER for HAS in the case of StyleTTS2 is 44.03\%, whereas for Neutral, it is 39\%. This indicates lower confidence of the anti-spoofing model in distinguishing between spoofed and bona-fide speech when it is exposed to fake emotional speech. These results reinforce our claim that SOTA anti-spoofing models, despite performing well on traditional datasets like ASVspoof 2019 and ASVspoof 2021, struggle with speech samples from recent emotionally expressive TTS models, especially when handling emotional speech, making them unreliable in such scenarios. Additionally, our analysis reveals that EER values vary across emotion categories; for instance StyleTTS2, the EER for Happiness is 45.00\%, for Sadness is 46.75\%, and for Anger is 35.17\%. 

\subsection{Finetuning RawNet2 on EmoSpoof-TTS}
Given the fact that, the pre-trained anti-spoofing method struggles to generalize to other datasets, we explored the performance when the model is trained with EmoSpoof-TTS. For this purpose, we fine-tuned the pretrained RawNet2 on the EmoSpoof-TTS using the train, validation and test subsets, which are described in Section 5.1. We structured the TTS models and speaker splits to ensure the training, validation, and test sets have no overlap, ensuring a fair evaluation. 
Results in Table \ref{tab:Performance_Emo-RawNet2} indicate a significant improvement in Emo-RawNet2's performance over the pre-trained RawNet2 model. For instance, with StyleTTS2, the EER for HAS in the pre-trained RawNet2 is 44.03\%, which drops to 10.36\% for Emo-RawNet2. However, performance disparities across emotions persist, as reflected in the EER values: 8.00\% for Happiness, 13.25\% for Sadness, and 5.83\% for the Neutral state. 

This uneven performance across different emotions suggests that attackers could exploit these inconsistencies by targeting emotions to which the model is more sensitive, increasing the likelihood of successful attacks. Moreover, the results in Table \ref{tab:Performance_Emo-RawNet2} highlights that while fine-tuning an existing model improves overall performance, it fails to resolve the disparity across emotions. Therefore, defending against the vulnerability of anti-spoofing models to emotionally expressive speech does not only require a useful dataset but also anti-spoofing paradigms specifically tailored for detecting emotional speech.


\renewcommand{\arraystretch}{1.3} 
\begin{table}[t!]
\centering
\caption{Performance of Pre-Trained RawNet2 model on EmoSpoof-TTS (EER\%)}
\label{tab:emospoof}
\scalebox{0.8}{
\begin{tabular}{c|c|c|c|c|c|c}
    \hline
    \multicolumn{7}{c}{\textbf{EmoSpoof-TTS}} \\ 
    \hline
    {\textbf{TTS Model}} & \textbf{HAS} & \textbf{Neutral} & \textbf{Happy} & \textbf{Angry} & \textbf{Sad} & \textbf{Overall} \\ 
    \hline
    \textbf{StyleTTS2} & \textbf{44.03} & 39.00 & 45.00 & 35.17 & 46.75 & 43.06 \\ 
    \textbf{F5-TTS} & \textbf{42.97} & 38.08 & 38.42 & 39.42 & 47.25 & 41.92 \\ 
    \textbf{CosyVoice} & \textbf{42.44} & 42.17 & 39.58 & 39.75 & 47.50 & 42.65 \\ 
    \hline
\end{tabular}}
\vspace{1mm}
\end{table}


\begin{table}[t!]
\centering
\caption{Performance of Emo-RawNet2 on the EmoSpoof-TTS test set (EER\%)}
\label{tab:Performance_Emo-RawNet2}
\scalebox{0.8}{
\begin{tabular}{c|c|c|c|c|c|c}
    \hline
    \multicolumn{7}{c}{\textbf{EmoSpoof-TTS (test set)}} \\ 
    \hline
    \textbf{Model} & \textbf{HAS} & \textbf{Neutral} & \textbf{Happy} & \textbf{Angry} & \textbf{Sad} & \textbf{Overall} \\ 
    \hline
    \textbf{\makecell{\textit{Emo-}\\\textit{RawNet2}}} & \textbf{10.36} & 5.83 & 8.00 & 9.25 & 13.67 & 9.65 \\ 
    \hline
\end{tabular}
}
\vspace{2mm}
\end{table}


\section{Proposed Method}

As discussed in Section 3, the performance of traditional anti-spoofing on emotional synthetic speech degrades significantly. Additionally, the uneven performance of anti-spoofing models across different emotions presents a security risk, as attackers may exploit the model’s vulnerability to specific emotional categories. To mitigate these vulnerabilities and enhance robustness against emotion-targeted attacks, this section introduces a novel approach: the Gated Ensemble Method (GEM).



\subsection{Emotion-specialized Anti-Spoofing models}

To build emotion-specialized anti-spoofing models, we use Emo-RawNet2, our general-purpose model described in section 3.2. We fine-tune Emo-RawNet2 on emotion-specific subsets of the EmoSpoof-TTS dataset, resulting in four emotion-specialized models: Model-H (Happiness), Model-A (Anger), Model-S (Sadness), and Model-N (Neutral state).


\subsection{Gated Ensemble Method (GEM)}

During inference, we lack prior knowledge of the emotion in the input speech, so we must automatically select the corresponding specialized anti-spoofing model. We chose not to use a hard selection of models, as we believe that emotions are inherently interconnected, rather than orthogonal \cite{northogonal}. To account for the relationships between different emotions, we propose incorporating contributions from multiple emotion-specialized models. This approach ensures a more holistic analysis, improving the model’s ability to detect whether a sample is fake. Therefore, we introduce an ensemble of specialized models with a soft gating mechanism to achieve the optimal solution.

The core idea of GEM is to ensemble predictions from specialized models in a way that gives the highest weight to the model associated with the emotion of the input signal 
$x$ during final decision-making. Soft gating enables this by allowing the most relevant model to make the final judgment while still incorporating the contributions from the other models.

For a given input $x$, each specialized model $M_i$ produces a score $S_i$:
\begin{equation}
S_i = M_i(x),
\end{equation}
where $S_i$ is the model's spoof prediction score for $x$.

 We employ a recent the SER system \cite{SER} to determine the emotion probabilities of $x$, as illustrated in Figure \ref{fig:figure1}$(b)$. The input $x$ is fed to SER, which outputs the probabilities ($E_i$) for each emotion category: \begin{equation} E_i = SER(x). 
 \end{equation}

The final decision score $y_s$ is computed as a weighted average of model scores, where weights are the emotion probabilities:
\begin{equation}
y_s = \sum_{i} S_i \cdot E_i.
\end{equation}

We control the softness of this weights by applying softmax with temperature $T$ to the output logits of the SER model. Based on $y_s$, the system determines whether the input signal is spoofed ($y=0$) or authentic ($y=1$).



\section{Experiments and Results}

\subsection{Dataset}




For the experiments in this paper, we partitioned the dataset into separate training, validation, and test sets based on speakers and TTS models to ensure a fair and unbiased evaluation:
\begin{itemize}
    \item \textbf{EmoSpoof-TTS Train set}: Includes 4 emotion categories and 4 speakers, with data generated by \textit{CosyVoice}, totaling 9,600 utterances (4,800 bona-fide and 4,800 spoofed).
    
    \item \textbf{EmoSpoof-TTS Validation set}: Includes 4 emotion categories and 2 speakers, with data generated by \textit{F5-TTS}, totaling 4,800 utterances (2,800 bona-fide and 2,800 spoofed).
    
    \item \textbf{EmoSpoof-TTS Test set}:  Includes 4 emotion categories and 4 speakers, with data generated by \textit{StyleTTS2}, totaling 9,600 utterances (4,800 bona-fide and 4,800 spoofed).
    
\end{itemize}
To obtain Emo-RawNet2, we fine-tune the pre-trained RawNet2 on the training and validation sets of EmoSpoof-TTS. For the emotion-specialized models, we extracted emotion-specific subsets from these sets and fine-tuned Emo-RawNet2 separately for each emotion.

\subsection{Implementation Details}

We utilize the pre-trained RawNet2 with default parameters\cite{RawNet2}. During the fine-tuning process to obtain the general-purpose model Emo-RawNet2, we used a batch size of 32 and trained the model for 50 epochs. For emotion-specific fine-tuning, the batch size was reduced to 8, and training was extended to 100 epochs. Learning rate was set to $ 1e^{-4}$ in both the fine-tuning process. SER \cite{SER} is trained to classify 4 emotion categories (Neutral, Happiness, Anger, Sadness) on the MSP-Podcast dataset \cite{msp-podcast}, using the default parameters and recipe provided. To ensure soft gating in the GEM, we set the temperature of the SER to 1.5.

\subsection{Results}

As a baseline, we consider the general-purpose model, Emo-RawNet2, which was fine-tuned with the full set of emotions from the EmoSpoof-TTS dataset, as described in Section 3.2. This allows us to analyze the effectiveness of our proposed method, which ensembles specialized models using a soft gating mechanism. Results in Table \ref{tab:Performance_Emo-RawNet2_GEM_EmoSpecific} presents the performance of our emotion-specialized models Model-N, Model-H, Model-A, Model-S. Each specialized model not only enhances detection for its respective target emotions but also demonstrates improvements in detecting other emotions. These results indicate the non-orthogonality \cite{northogonal} of emotions, as improving the model's performance on a particular emotion affects others. 

\begin{table}[!t]
\centering
\caption{Performance Comparison on EmoSpoof-TTS test set (EER\%)}
\label{tab:Performance_Emo-RawNet2_GEM_EmoSpecific}

\scalebox{0.8}{
\begin{tabular}{c|c|c|c|c|c|c}
    \hline
    & \multicolumn{6}{c}{\textbf{EmoSpoof-TTS (test set)}} \\ 
    \hline
    \textbf{Model} & \textbf{HAS} & \textbf{Neutral} & \textbf{Happy} & \textbf{Angry} & \textbf{Sad} & \textbf{Overall} \\ 
    \hline
    \textbf{Model-N} & 7.89 & \textbf{3.50} & 7.92 & 7.67 & 7.08 & 6.79 \\ 
    \textbf{Model-H} & 6.03 & 2.08 & \textbf{6.67} & 4.00 & 5.75 & 5.02 \\ 
    \textbf{Model-A} & 11.36 & 9.58 & 13.25 & \textbf{7.08} & 12.83 & 11.02 \\ 
    \textbf{Model-S} & 7.89 & 3.58 & 9.00 & 5.08 & \textbf{7.67} & 6.92 \\ 
    \hline
    \textbf{GEM} & \textbf{6.22} & 2.92 & 6.75 & 5.58 & 5.75 & 5.33 \\ 
    \hline
\end{tabular}
}
\vspace{2mm}
\end{table}



 


The results presented in Table \ref{tab:Performance_Emo-RawNet2_GEM_EmoSpecific} demonstrate that GEM not only enhances the overall performance of the anti-spoofing model compared to Emo-RawNet2 but also mitigates the imbalance in spoof detection across different emotions. Its performance is notably more consistent across emotions compared to Emo-RawNet2 and individual emotion-specialized models. The results shows that proposed method (GEM) comes very close to the optimal performance of each specialzed model for its respective emotion category leading to a much improved overall performance and more balanced performance across different emotions. Another key strength of GEM is that it enhances spoof detection in emotional speech without compromising performance on neutral speech. The above results demonstrate that handling spoof detection for all emotions together is a complex task for a single anti-spoofing model, resulting in suboptimal performance. In contrast, using multiple models specialized for each emotion not only improves emotion understanding but also captures inter-emotional relationships, leading to better overall anti-spoofing performance.


\section{Conclusion}
In this work we explored the effect of emotion on anti-spoofing, revealing emotion-related risks and the need of emotional aspect for this task in terms of both dataset and methodology. Our analysis showed current anti-spoofing paradigm with datasets and methods without emotional focus struggle with emotion-targeted attacks. 
To facilitate research in this domain, firstly we introduced EmoSpoof-TTS, a corpus created using recent text-to-speech (TTS) models to benchmark and enhance anti-spoofing performance on emotional synthetic speech. Furthermore, we proposed an anti-spoofing method specifically tailored for tackling emotional aspect of anti-spoofing. The proposed method, the Gated Ensemble Method (GEM), which leverages emotion-specialized models and a soft gating mechanism, improved anti-spoofing performance on emotional speech, validating the need for such methods. For future work, we aim to enhance anti-spoofing resilience to emotion-targeted attacks across diverse emotional states and TTS models while expanding EmoSpoof-TTS.

\section{Acknowledgment}
This work is supported by NSF CAREER award IIS-2338979.

\bibliographystyle{IEEEtran}
\bibliography{mybib}

\end{document}